# A heuristic algorithm for straight-line embedding of a hamiltonian cycle onto a given set of points inside simple polygons

Maryam Fadavian[1] , Heidar Fadavian[2]

[1] Department of Computer Engineering, Amirkabir University of Technology, Tehran, Iran
[2] Department of Electrical and Computer Engineering, Tarbiat Modares University, Tehran, Iran

**Abstract:** This paper investigated the problem of embedding a simple Hamiltonian Cycle with n vertices on n points inside a simple polygon. This problem seeks to embed a straight-line cycle (without bends), which does not intersect either itself or the sides of the polygon, *i.e.*, it is planar. This problem is a special case of an open problem to find a simple Hamiltonian (s, X, t)-path (a simple path that starts at s and ends at t, where s, t, and all other vertices within the path are a member of set X) inside a simple polygon, which does not intersect itself or the sides of the polygon. The complexity of the problem in this paper is not verified yet, and it is an open problem. However, similar problems are resolved that are NP-Complete. A heuristic algorithm with time complexity of $O(r(n^2m + n^3))$ and space complexity of $O(n^2 + m)$ is proposed to solve the problem.

## 1 Introduction

The geometric embedding of the graph is one of the well-known problems regarding graph theory and computational geometry. This problem constituents a graph comprising n nodes and a set of n points on a plane. It aims to find a one-to-one mapping between vertices of the graph and the set of points such that several constraints are observed and some quantities are optimized. For instance, one of the limitations might be the planar embedding, meaning that the edges of the embedded graph must not intersect each other, or the edges must be straight-line [1]. Most studies in this regard embed the graph on the points of the plane to shorten the length of the edges of the embedded graph, which is quite helpful in designing the distributed computer network (DCN) [2]. However, some studies have considered the constraint of the point enclosing polygon and carried out embedding inside the polygon. When considering this constraint, for planar embedding, the embedded graph edges should neither intersect each other nor the sides of the polygon. The graph might not be embedded as a straight-line planar on the points provided inside the polygon. Thus, there might be intersecting in some cases, or the embedding of some edges might not be straight-line. In other words, an edge might be

🖂 Heidar Fadavian
E-Mail: heidarfadavian@modares.ac.ir

Maryam Fadavian
E-Mail: m.fadavian@aut.ac.ir

embedded as a multilinear. The place of breakage during multilinear edge embedding is called a bend. Thus, in this case, the number of edge bends is one of the quantities that can be minimized.

The graph embedding problem is quite useful. For instance, the well-known Euclidean Traveling Salesman Problem can be considered a special case of the problem of embedding a graph on a set of points such that the sum of the Euclidean lengths of the embedded edges should be minimized. Another famous problem is the median calculation problem for a star graph as the common topology of a distributed network. In some applications, the graph is embedded on a plane constrained with simple polygons. For example, consider a graphical user interface that is supposed to display a graph in a star-shaped polygon or the PCB design of an electronic device, which must be U-shaped [3]. Its other application is in the field of robotics. A robot is trying to move from the source toward the destination by passing through a specific number of stations. The movement route must be found for the robot to have the minimum rotation and direction change. Designing electrical circuits is another application of this problem in which places are specified for placing the components of the circuit, and routing should be carried out such that the connections have the minimum bend.

This paper examined the straight-line planar embedding of a Hamiltonian Cycle on the specified points inside a simple polygon. Having a simple general polygon Q with m vertices and set X with n points inside that polygon, the simple Hamilton cycle P with a length of n is embedded on points of X such that the P vertices are placed on the points of X. P does not intersect either its edges or the sides of Q polygon. At the same time, the edges of the cycle are straight lines (without bends). In this problem, the number of P vertices is the same as the number of points X and is equal to n. Consequently, all n points of set X must be used, and no auxiliary points can be used, *i.e.*, all edges of the cycle are straight-line. In addition, a simple path, cycle, or polygon means that they do not intersect each other's edges or sides.

Section 2 reviews the related literature. The proposed heuristic algorithm to solve the discussed problem is introduced in Section 3. Section 4 includes the assessment and results of experimental tests of the proposed algorithm. Finally, the conclusion and a discussion on the open problems in this field are given in Section 5.

## 2. Literature Review

Several algorithms are provided for embedding graphs on indefinite space to minimize the length of edges. They are provided in finite space to minimize edges bend. Gritzmann *et al.* [4] proved that there is a straight-line graph embedding on any set of points, only if that graph is outer-planar.

Alsuwaiyel and Lee [5] proved that finding a Hamilton (s, X, t)-path, which is not necessarily a simple path inside a simple polygon, is an NP-Complete problem. This proof stands correct when the point set of X is limited to polygon points. Remember that polygon boundary is not a solution since s and t are not necessarily consecutive. Bose *et al.* [6] proposed an optimal algorithm for the planar straight-line embedding of the tree on the set of points. Cheng *et al.* [7] proved that finding a simple (s, X, t)-path, despite desired obstacles (not a simple polygon), is an NP-Complete problem in case it does not encounter any obstacle. Bose [8] presented an algorithm with the time complexity of $O(n\log^3 n)$ and space complexity of O(n) for the planar straight-line embedding of the outer-planar graph on the set of points. Cabello [9] proved that the decision-making problem regarding the straight-line embedding of a planar graph on a set of points is an NP-Complete problem.

Di Giacomo *et al.* [10] provided a heuristic algorithm for embedding a tree on the points inside a simple polygon. The proposed algorithm ensures that no edge has more than r/2 bends, where r is the number of non-convex vertices of the polygon. Binucci *et al.* [11] examined the problem of computing the straight-line ascending embedding of a directed acyclic planar graph G into a set of points S.

Sepehri and Bagheri [12] proposed an algorithm to embed a tree with n vertices on n points inside an m-vertex polygon to minimize the number of bends of the obtained tree. Nishat *et al.* [13] suggested an algorithm that determines the existence or absence of the straight-line embedding of planar 3-tree G with n vertices on set S with n points. It achieves such embedding when it exists. Noorbakhsh and Bagheri [14] provided two heuristic algorithms to solve the problem of tree embedding on a set of points on the plane with the minimum length such that the number of points inside the plane is the same as the nodes of the provided tree. Furthermore, the embedded tree is isomorphic with the main tree. Norouzitallab and Bagheri [1] suggested several heuristic algorithms for embedding trees on a set of points with the minimum total length. These include heuristic algorithms based on the traveling salesman problem, the nearest neighbors, based on the largest common forest, clustering-based, degree-based partitioning, improvement of the algorithm based on clustering and last optimization. According to the experimental results, the degree-based partitioning algorithm outperformed the other algorithms. Tan and Jiang [15] proposed an algorithm with a time complexity of $O(\log m (n^2 + m))$ and space complexity of $O(n^2 + m)$ for embedding an edge on set X of n points inside a simple m-vertex polygon Q. The vertices of the beginning and end of the edge must be mapped on two specific points $s, t \in X$. The embedded edge should be either a straight line or has the minimum bend. These bends are merely placed on point X. Moreover, the embedded edge should not intersect the boundaries of Q and itself. This algorithm improved the two former algorithms [7 and 16], one of which with time and space complexities of $O(m^2n^2)$ and the other with time complexity $O(n^3\log m + mn)$ and space complexity of $O(n^3 + m)$. Razzazi and Sepahvand [17] proved the NP-Completeness of simultaneous embedding of two separate simple paths, in which vertices of one path are red, and all vertices of the other path are blue. In other words, the red vertices are embedded on red points while the blue vertices are embedded on blue points. Furthermore, they proposed a heuristic algorithm with a mean of time order $O(n^3)$ to solve this problem. They also proved that finding a simple Hamilton (s, X, t)-path, with the presence of desired obstacles such that the obtained path does not encounter them, is an NP-Complete problem.

Fadavian and Fadavian [18] suggested a genetic algorithm (GA)-based metaheuristic algorithm for the problem addressed in this paper. The proposed GA is applicable to solve this problem by providing an appropriate mutation operator. According to the experimental results, the proposed GA based on this mutation operator enjoys a higher quality than the two-point ordinary mutation operator.

## 3. Heuristic Algorithm

This paper investigated the problem of the straight-line planar cycle embedding on a set of points inside a simple polygon. Considering that the complexity of this problem is not verified yet and it is an open problem, a heuristic algorithm has been proposed to solve it. The idea of this algorithm is first to partition the problem input, a simple polygon, into several convex polygons through one of the polygon partitioning algorithms called Hertel–Melhorn. The dual graph of partitioning is then calculated. The obtained dual graph is always a tree since partitioning is based on the diagonals drawing. Therefore, its Euler tour can be calculated. A straight-line planar cycle is embedded on the points inside each of the convex polygons. Following the obtained Euler tour, two non-intersecting adjacent connection edges should be found between the two adjacent convex polygons. Finally, the straight-line planar cycle embedding on the points inside the provided simple polygon is achieved by embedding the connection edges and eliminating the edges between two connection vertices in the convex polygons. The states differ depending on whether the convex polygons have zero points, one point, two points, or more, and this algorithm covers all states as far as possible. Furthermore, initiatives were made to choose connection vertices so that the final cycle is a planar and straight-line result.

### 3.1. Heuristic Algorithm Pseudocode

It is assumed that the points inside a simple polygon are all at a general position for straight-line planar embedding. In this pseudocode, the polygon Q is partitioned into several convex pieces at line 1. In line 2, the dual graph of this partitioning is calculated using DCEL. Considering that partitioning is based on the diagonals drawing, the obtained dual graph is a tree. In line 3, the dual graph is examined to see whether the presented simple polygon is a convex or a concave. In the case of a convex polygon, it can be ensured that a straight-line planar cycle can be embedded on the points inside it. After that, one of the available algorithms is introduced in this regard.

If it is a concave polygon, the Euler tour of the dual partitioning tree is calculated in line 4.a. Each node in the dual tree equals a convex piece from the partitioning. In line 4.b, the provided points inside the simple polygon are partitioned into the set of points inside each convex pieces. As we know the number of convex pieces depends on concave vertices (r). Consequently, there will be sets of distinguished points equal to the number of the convex pieces, whose union will result in a set of input points, and their sharing will be null. In line 4.c, a straight-line planar cycle can be embedded on the set of points inside each convex pieces if it is not null (with the algorithm presented to generate a simple polygon). If the set of points comprises one point, there will be no cycle, and in case it comprises two points, a straight-line edge will be embedded.

In line 4.e, the Eulerian tour traversing will begin to search for a connection edge between each two simple polygons, inside each two convex pieces that their peer nodes are observed in the Eulerian tour until returning to the start point. The ComputeConnectionEdge ($p_1$, $n_1$, $p_2$, $n_2$) will be called to find the connection edge between two simple polygons. The pseudocode of this function is provided herein-below. During the Eulerian tour traversing, in case there are nodes whose peer convex piece lacks any points, the next nodes of the tour will meet in the order of traversing to reach the first nodes, whose peer convex piece comprises at least one point. Then, we will search for the connection edge with a simple polygon inside the peer convex piece of this node.

Finally, after the completion of the Eulerian tour traversing and finding the connection edges at line 4.g, in each two simple polygons (including at least three vertices) which there are two non-intersecting adjacent connection edges between them, the side whose two ends are the connection vertices to connect with the other polygon will be eliminated. Finally, a straight-ling planar cycle will be achieved on the points inside the given simple polygon.

**The proposed algorithm for planar straight-line embedding of a cycle**

**Input:** simple m-polygon Q, set X of n points inside Q
**Output:** the embedding of the cycle on the points of X

1. Partition Q into Convex Pieces
2. Compute the Dual Graph of Partitioning
3. if the Dual Graph is a Node

a. "The Simple Polygon is a Convex Polygon, so Always Can Embed a Simple Geometric (planar and straight-line) Cycle on The Points inside The Convex Polygon"

4. else
    a. Compute the Eulerian-tour of the Dual Graph
    b. Compute the Set of Points of each Convex Pieces ($X_i$ ($0 \leq i \leq r+1$))
    c. for each Set of Points ($X_i$):
        i. if $k_i$ (the number of points of $X_i$, $0 \leq k_i \leq n$) != 0:
            1. SimplePolygonGeneration($X_i$, $k_i$)
        ii. endif
    d. endfor
    e. while the Traversing of the Eulerian-tour is not finished ($j$ < Length of Eulerian-tour), do the followings:
        i. if $k_j$ != 0 and $k_{j+1}$ != 0
            1. ComputeConnectionEdge($C_j$, $k_j$, $C_{j+1}$, $K_{j+1}$)
        ii. elseif $k_j = 0$
            1. $j = j+1$
            2. ComputeConnectionEdge($C_j$, $k_j$, $C_{j+w}$, $K_{j+w}$) such that $k_{j+w}$ != 0 and $j < j+w <$ Length of Eulerian-tour and there is not exist $k_{j+p}$ such that $k_{j+p}$ != 0 and $p < w$
        iii. elseif $k_{j+1} = 0$
            1. ComputeConnectionEdge($C_j$, $k_j$, $C_{j+w}$, $K_{j+w}$) such that $k_{j+w}$ != 0 and $j < j+w <$ Length of Eulerian-tour and $C_j$ != $C_{j+w}$ and there is not exist $k_{j+p}$ such that $k_{j+p}$ != 0 and $p < w$
        iv. endif
    f. endwhile
    g. for each two ConnectionEdges between two polygons:
        i. if the number of each polygon vertices are more than two:
            1. Remove the edge between connection vertices of each polygons
        ii. endif
    h. endfor
5. endif

### 3.2. Connection Edge Computation Algorithm

An algorithm is suggested to calculate the connection edge between two simple polygons. The proposed algorithm receives two polygons with the number of their vertices as inputs. If a connection edge is available, the algorithm will restore it as the connection edge between these two polygons. A connection edge does not intersect simple polygons Q, $p_1$, and $p_2$; it intersects none of the available connection edges.

When calculating the connection edge between two polygons, the algorithm first examines whether there was a connection edge between these two polygons, then operates based on these two states. Suppose it is the first time that it wants to find a connection edge between two polygons given in the input, it calculates the distance between all the vertices of the given polygons using Relation 1 and restores them together with the two involved vertices.

$$\sqrt{(v_1 \cdot x - v_2 \cdot x)^2 + (v_1 \cdot y - v_2 \cdot y)^2} \tag{1}$$

where $v_1$ and $v_2$ are vertices from $p_1$ and $p_2$, respectively.

Then, the list obtained from a distance will be ordered ascending. From the beginning of the list that the two vertices are at their shortest distance, it commences investigating the required conditions for the connection edge. Once the first edge is found, its connection will be restored as the connection edge between those two polygons, ending the investigation. When it reaches the end of the list without finding any connection edge, it restores -1 at the place of maintaining the connection edge between these two polygons and ends the algorithm so that it will not have to look for a connection edge between these two polygons.

However, if a connection edge has already been calculated between these two given polygons at the input, it checks to see its value is not -1 since the algorithm will end in this case. Suppose polygon $p_2$ has merely one vertex with two connection edges passing through it. In that case, it cannot be selected as the connection vertex, and polygon $p_1$ must search for another polygon for connection. Consequently, it continues the Eulerian tour traversing to meet a node whose peer simple polygon inside the convex segment does not have only one vertex; if it does, a maximum of one connection edge has passed through it. Once that node is found, it calculates and restores the distance between all $p_1$ vertices and the found polygon that meets the conditions, except for connection vertices in the polygons (*i.e.*, when it comprises merely one point, that point is authorized). Then, it continues the said routine to find the connection edge.

Suppose none of the above states are established, meaning that, there is one connection edge between these two given polygons, and it is already calculated and restored. In that case, the algorithm calculates the vertices near the connection vertex. If $p_1$ or $p_2$ merely have one point, *i.e.*, have no adjacent vertices, they are regarded as their adjacent vertices. Then, all possible edges might be calculated from the connection of the adjacent vertices to each other. Considering that each vertex has a maximum of two adjacent vertices, a maximum of four edges can be selected as the connection edge between these two polygons. Then, the algorithm examines all these candidate edges, looking for the conditions of the connection edge. Once it has found the first connection edge, it will restore it as the connection edge between those two polygons and ends the investigation.

When none of the selected edges met the conditions of the connection edge (the candidate edge might meet all the conditions and merely intersect its adjacent connection edge), for each candidate edge, the algorithm displaces the connection order of the candidate edge vertices and the previously found connection edge vertices. Then, it examines the two newly found edges to ensure whether they met the conditions of the connection edge. Immediately after the first two non-intersecting adjacent connection edges, it restores them as connection edges between those two polygons and eliminates the previous connection edge, and then, the algorithm ends. If it fails to find two non-intersecting adjacent connection edges after the displacement, the algorithm ends and will restore the same connection edge in its memory found earlier.

**Algorithm: ComputeConnectionEdge($p_1$, $n_1$, $p_2$, $n_2$)**

**Input:** let $p_1$, $p_2$ are polygons and $n_1$, $n_2$ are the number of vertices of polygons in order
**Output:** A ConnectionEdge from $p_1$ to $p_2$, if there exist

1. if not already computed the ComputeConnectionEdge($p_2$, $n_2$, $p_1$, $n_1$)
    a. Compute and store the distance between all the pairs of vertices of polygons $p_1$ and $p_2$

b. Sort the list of distances in increasing order
c. for each element of distanceslist:
    i. if its segment does not intersect none of ConnectionEdges, $p_1$, $p_2$ and Q
        1. Store it as a ConnectionEdge between $p_1$ and $p_2$
        2. break
    ii. endif
d. endfor
e. if there is not exist a ConnectionEdge between $p_1$ and $p_2$
    i. Store -1
f. endif

2. else
    a. if there is not exist a ConnectionEdge between $p_2$ and $p_1$
        i. break
    b. elseif $n_2 = 1$ and $p_2$ has two ConnectionEdge
        i. Continue to traversing Eulerian-tour to find a convex piece such that $n_i$ != 1 or $p_i$ has at most one ConnectionEdge
        ii. Compute and store the distance between all the pairs of vertices of polygons $p_1$ and $p_i$ except the vertices that have ConnectionEdges if the number of polygon vertices are more than one
        iii. Sort the list of distances in increasing order
        iv. for each element of distanceslist:
            1. if its segment does not intersect none of ConnectionEdges, $p_1$, $p_i$ and Q
                a. Store it as a ConnectionEdge between $p_1$ and $p_i$
                b. break
            2. endif
        v. endfor
    c. else
        i. Compute all the adjacent vertices of the vertices of the ConnectionEdge between $p_2$ and $p_1$ (if $n_1 = 1$or $n_2 = 1$, consider itself instead of its adjacent vertex)
        ii. for each segment that created from adjacent vertices:
            1. if it does not intersect none of ConnectionEdges, $p_1$, $p_2$ and Q
                a. Store it as a ConnectionEdge between $p_1$ and $p_2$
                b. break
            2. endif
        iii. endfor
        iv. if none of the segments do not select as a ConnectionEdge between $p_1$ and $p_2$
            1. for each segment:
                a. Change the order of vertices connection of segment and ConnectionEdge between $p_2$ and $p_1$ and create two new segments
                b. if each new segments do not intersect none of ConnectionEdges, $p_1$, $p_2$ and Q
                    i. Remove the ConnectionEdge between $p_2$ and $p_1$

ii. Store the new segments as ConnectionEdges between $p_1$ and $p_2$
iii. break
c. endif
2. endfor
v. endif
d. endif
3. endif

### 3.3. Complexity Algorithm of Connection Edge Computation

1. The distance between all vertices of polygons $p_1$ and $p_2$ can be calculated in $O(n_1.n_2)$ time.
2. 2.The list of distances can be sorted in $O(n_1 n_2 \log n_1 n_2)$ time.
3. The intersection of one edge with other connection edges, polygons $p_1$, $p_2$, and Q can be calculated in $O(n_e)$, $O(n_1)$, $O(n_2)$, and $O(m)$ times, respectively. These intersections will be checked for a maximum of $O(n_1.n_2)$ times to find the first connection edge ($n_e$ and m equal the number of edges of the dual tree and the number of the vertices of polygon Q, respectively).
4. The adjacent vertices of each vertex of polygons $p_1$ and $p_2$ can be calculated when the order of the connection of vertices to each other is available in $O(n_1)$ and $O(n_2)$ times, respectively. Each vertex has a maximum of two adjacent vertices merely when there is a polygon, *i.e.*, the number of vertices exceeds two.

Consequently, the time and space complexities of the total algorithm equal $O(n_1 n_2 m + n_1 n_2^2 + n_1^2 n_2)$ and $O(n^2)$, respectively.

### 3.4. Generation of a Simple Polygon

Embeding a straight-line planar cycle on a set of given points inside a convex polygon, or in general, on a 2D plane, will be the same as generation of a simple polygon on that set of points passing all points. Most algorithms are concerned with generation of simple polygons, one of which is explained below.

**Algorithm: SimplePolygonGeneration(S, n)**

**Input:** let $S = \{p_1, p_2, \ldots, p_n\}$ and n = number of points.

**Output:** $C = \{q_1, q_2, \ldots, q_n\}$ shows the simple polygon.

1. $C \leftarrow$ null
2. if $n \leq 2$
   a. Add the points to C and return C.
3. else
   a. Find the leftmost point ($p_i$ ($1 \leq i \leq n$)).
   b. Find the rightmost point ($p_j$ ($1 \leq j \leq n$)).
   c. Partition the points into A, the set of points below $p_i p_j$, and B, the set of points above $p_i p_j$ [you can use the left turn test on ($p_i$, $p_j$, ?) to determine if a point is above the line].

---

   d. Sort A by x-coordinate (increasing).
   e. Sort B by x-coordinate (decreasing).
   f. Return the polygon C defined by $p_i$, the points in A, in order, $p_j$, the points of B in order.

4. endif

---

### 3.5. Time Complexity of Algorithm for Generation of a Simple Polygon

1. The leftmost and rightmost point can be found in O(n) time.
2. Points can be partitioned based on whether they are above or below the line pipj in O(n) time.
3. Points can be sorted based on any component either in ascending or descending order in O(nlogn) time.
4. The points can be listed in the order of their connection to each other in O(n) time.
5. Consequently, the whole algorithm has a time complexity of O(nlogn).

### 3.6. Proving the Accuracy of Algorithm for Generation of a Simple Polygon

The accuracy of an algorithm can be proved using constructive proof. All points around $p_i$ and $p_j$ are placed in sets A or B. Therefore, the polygon obtained from line 3.f is generated using all points. It should be proved that none of the obtained sides of the polygon intersect another.

Consider each side of the output polygon. The first edge from $p_i$ to the first point in set A cannot intersect any segments (since there is no segment yet). Considering that the points of set A are processed based on its coordinate X, the next segment will be on the right side of the previous segment, and all previous segments will be on its left side. Consequently, no intersection occurs when it starts from $p_i$ and processes all points in set A to reach point $p_j$.

The same reasoning applies when returning from $p_j$ and processing all points of set B based on their coordinate X. Consequently, the achieved segments do not intersect each other. These segments intersect none of the segments obtained from set A since the points of set A are located below the line $p_i p_j$, and the points of set B are located above this line. Therefore, none of the sides intersects, and the generated polygon is simple.

### 3.7. Complexity of Heuristic Algorithm

1. Partitioning the given simple polygon Q into several convex pieces via the Hertel-Melhorn algorithm has a time and space complexity of O(m) (m is the number of the vertices of the simple polygon Q). Even though this algorithm obtains up to a maximum of four times the number of the optimal convex pieces, it is quite better in practice and obtains the optimal partitioning in most cases.
2. The dual partitioning graph can be calculated using its DCEL in O(r) time. According to Lemma 2-5-2 [19], each concave vertex requires a maximum of two necessary diagonals. Therefore, the number of necessary diagonals cannot exceed 2r, where r is the number of concave vertices (less than that might indicate that some diagonals are necessary for both end vertices). $\vec{e}$‹ IncidentFace($\vec{e}$), and IncidentFace(Twin($\vec{e}$)) are calculated per necessary diagonals to find the adjacent convex pieces (each diagonal is limited to two convex pieces). As a result, their peer nodes in the dual graph will be connected via an edge. Furthermore, it is known that the number of concave vertices (r) in a simple m-vertex polygon equals a maximum of m-3, *i.e.*, r = O(m).

3. The Euler tour of the dual graph, a tree, can be calculated using the linked list in the linear time in terms of the number of the edges of the tree ($n_e$). The number of the tree's edges equals the number of the partitioning diagonals, *i.e.*, O(r).
4. A set of points inside each convex piece can be calculated in O(nr) time.
5. The straight-line planar cycle inside all sets of points can be calculated in (r.nlogn) time. Since the number of the sets of points amounts to that of the convex pieces equal to O(r), the SimplePolygonGeneration(S, n) algorithm with a time order of O(nlogn) can be called for a maximum O(r) times.
6. The Euler tour length equals $2n_e$+1, where $n_e$ is the number of tree edges. Consequently, the ComputeConnectionEdge($p_1$, $n_1$, $p_2$, $n_2$) algorithm is called O($n_e$) times.
7. The number of connection edges equals O($n_e$) at its best, in which two non-intersecting adjacent connection edges exist between each two adjacent simple polygons. So it equals 2(2$n_e$), precisely.

As a result, the time complexity of the whole algorithm equals $O(r(n^2m + n^3))$. Furthermore, the complexity of the consumed memory in this algorithm equals $O(n^2 + m)$ on account of using the hash table.

### 3.8. The Special State of the Algorithm

As already mentioned, the proposed heuristic algorithm calculates the distance between all vertices of two polygons to find the connection edge between two simple polygons. It orders the obtained list increasingly, and once it has found the first connection edge, it will end. However, other states are possible. If it proceeds based on the shortest distance, even though the problem will lead to a solution, it cannot be obtained. Figure 1 is an instance of these states. Depending on the position of points inside each convex piece, the proposed algorithm might find a solution. The red lines in this figure are the necessary diagonals for partitioning, and the blue lines are the connection edges. Figure 1(A) provides an example where the problem has a solution, but the algorithm cannot find it. Figure 1(B) shows the correct solution in Figure 1(A). Figure 1(C) provides an example where the problem has a solution, and the heuristic algorithm can find it.

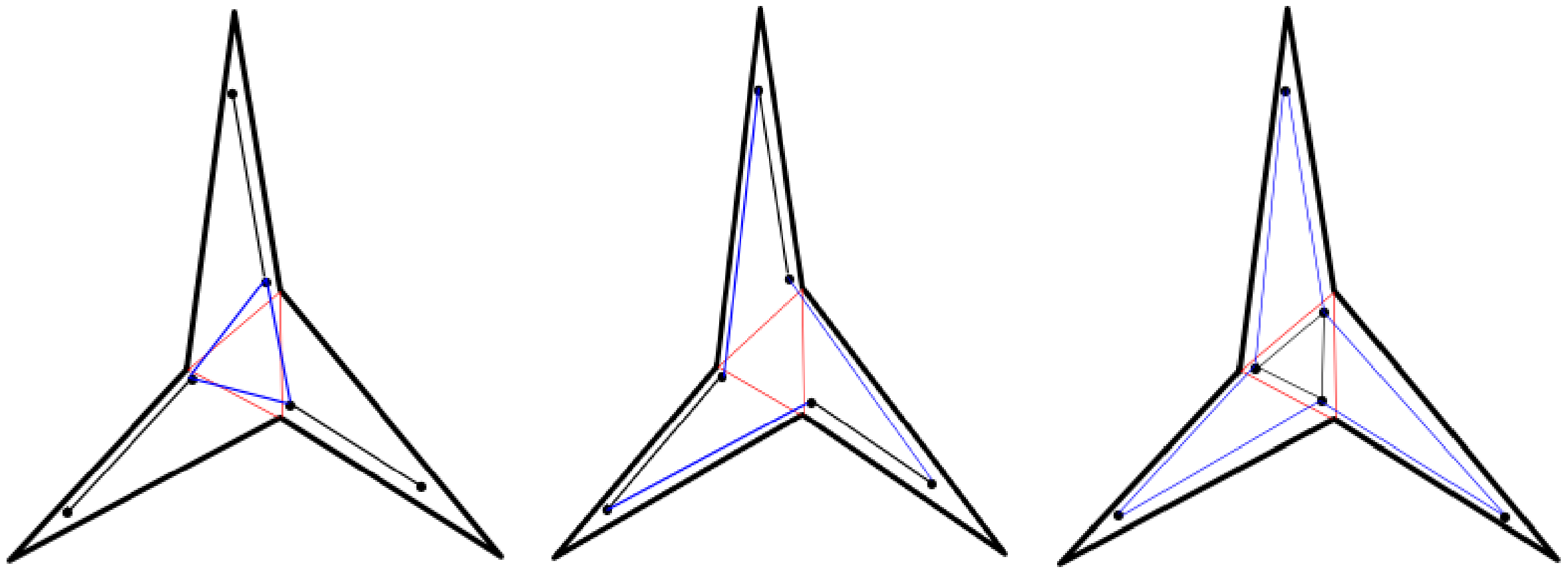

**(A)** **(B)** **(C)**

**Figure 1:** The Special State of the Algorithm

## 4. Assessment and Experimental Results of the Proposed Algorithm

This section examines the results of executing the heuristic algorithm and the mean of its number of successes in different states. Finally, samples of executing the heuristic algorithm are demonstrated. Several parameters are considered to assess the behaviors of the algorithms. For instance, to measure the impact of the number of sides on the algorithm's behavior, the problem was tested on various polygons with a number of sides (5-Gon, 10-Gon, 15-Gon, 20-Gon, and 25-Gon). Twenty-five different polygons were considered for each number of sides at the same number of sides to observe the impact of the polygon shape on the behavior of the algorithm, and the mean of the results was obtained. In addition, the number of points of set X inside the polygon was considered different (5 points, 10 points, 15 points, 20 points, 25 points, and 30 points). The proposed algorithm is implemented in Python in the JetBrains PyCharm IDE.

### 4.1. Polygons with Different Number of Sides on Fixed Points

Figure 2 shows the mean number of successes of the heuristic algorithm in finding the problem's solution per polygons with a different number of sides (5-Gon, 10-Gon, 15-Gon, 20-Gon, and 25-Gon) on the fixed number of points. This figure shows the output for 20 points inside each polygon.

Figure 2 demonstrates that adding the number of sides of a polygon decreases the mean number of successes because, in most cases, increasing the number of sides of a polygon increases the number of its concave vertices and convex pieces. However, considering that the number of points is fixed, the possibility of two connection edges decreases between each two adjacent convex pieces, and the problem had a solution (through establishing a connection edge between non-adjacent convex pieces).

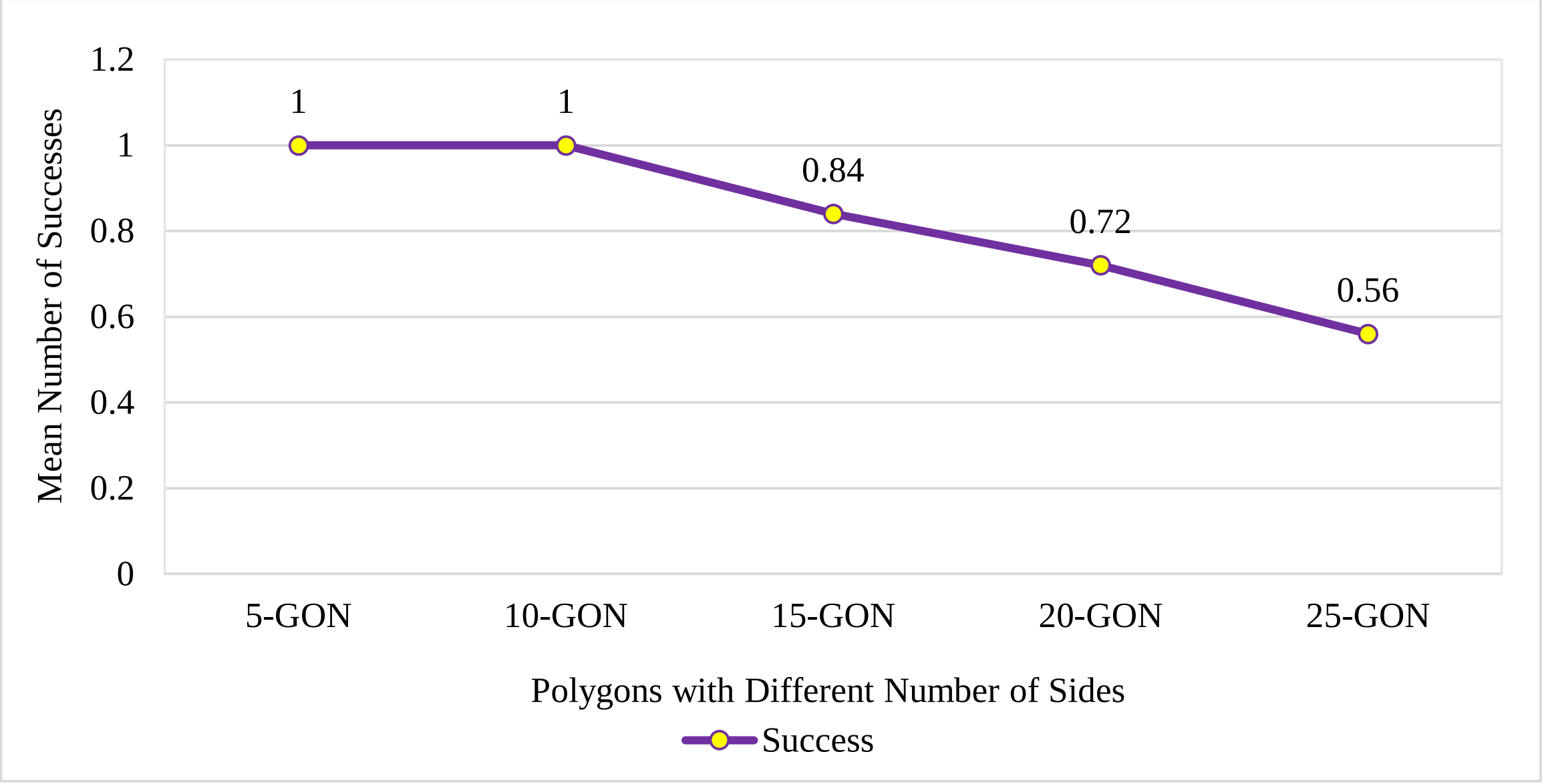

**Figure 2:** Mean Number of Successes per the Polygons with Different Number of Sides on 20 Points

### 4.2. Number of Different Points on the Fixed Polygons

Figure 3 shows the mean number of successes of the heuristic algorithm in finding the problem solution per the number of different points (5 points, 10 points, 15 points, 20 points, 25 points, and 30 points) on polygons with a fixed number of sides. This figure manifests the output for the 20-Gon.

Figure 3 indicates that the mean number of successes has increased by adding the number of points. With increasing the number of points inside polygons with a fixed number of sides, the possibility of two connection edges between two adjacent convex pieces increases, and the algorithm will be quite successful.

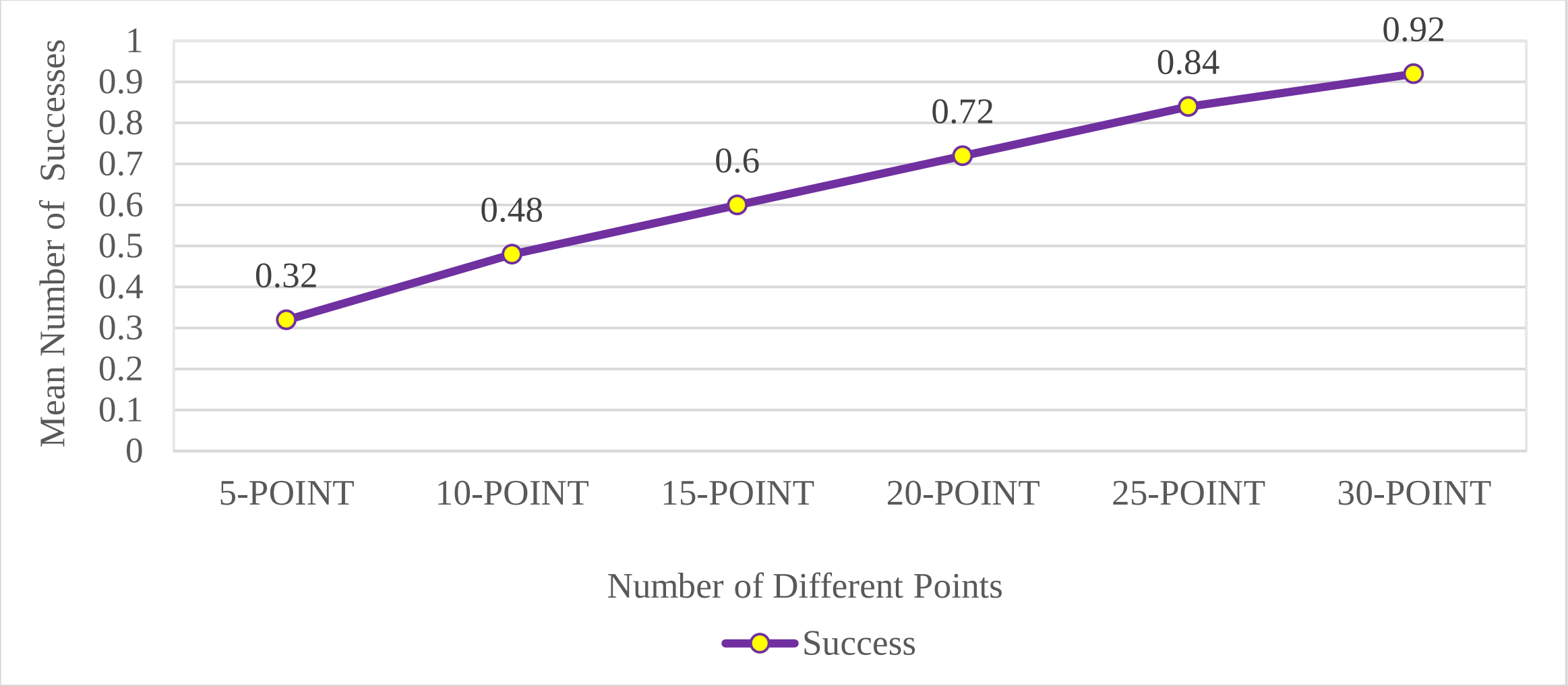

**Figure 3:** Mean Number of Successes per the Number of Different Points on a Fixed 20-Gon

### 4.3. Polygons with Different Number of Sides on Different Number of Points

Figure 4 shows the mean number of successes of the heuristic algorithm in finding the problem solution per polygons with a different number of sides on a different number of points. Considering the results demonstrated in this figure, per the specific number of sides, a test was carried out on several polygons with the same number of sides and on a different number of points. Then, the mean of the results is obtained.

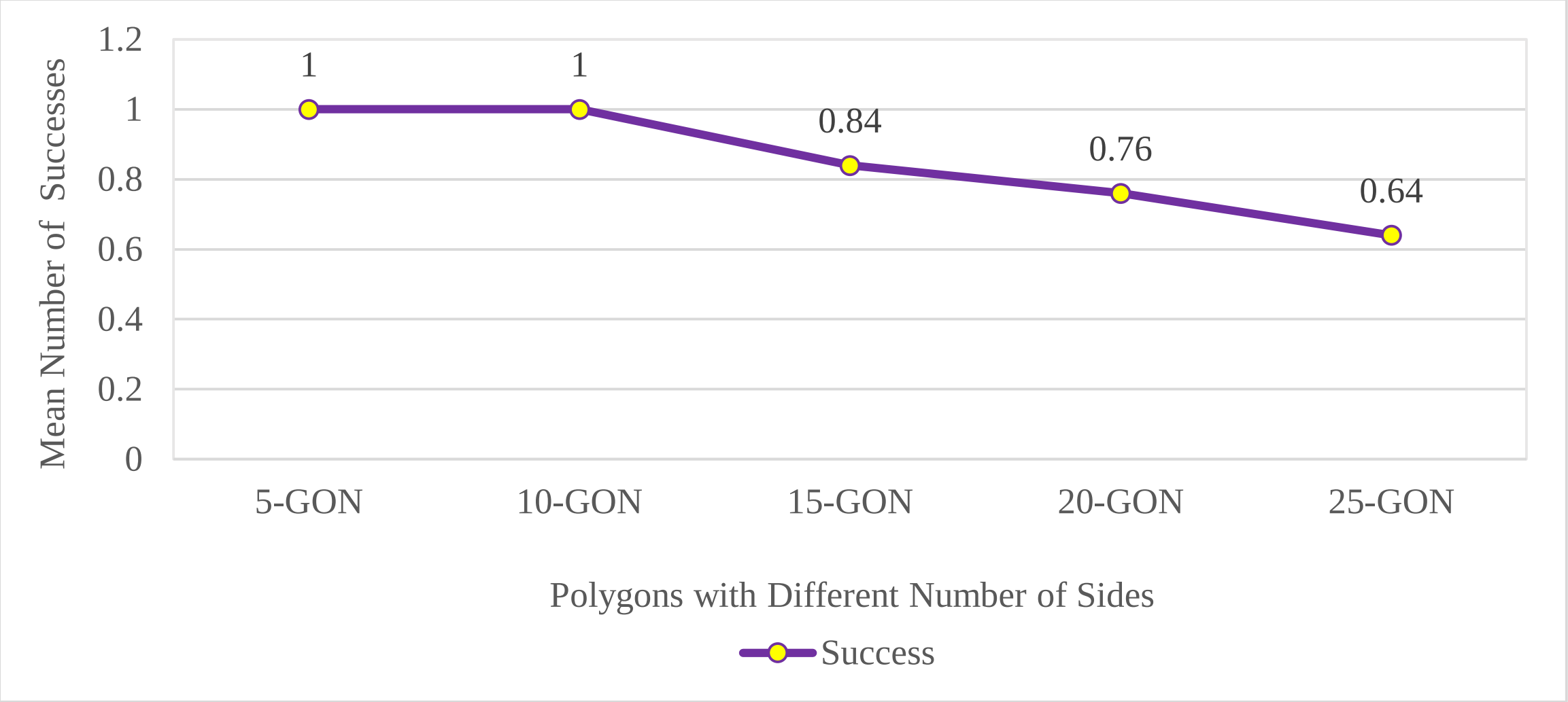

**Figure 4:** Mean Number of Successes per the Polygons with Different Number of Sides on Different Number of Points

### 4.4. Set of Points with a Different Number of Points on Different Polygons

Figure 5 shows the mean number of successes of the heuristic algorithm in finding the problem solution per different number of points on polygons with a different number of sides. It signifies that, considering the results in this figure, per the specific number of points, a test was carried out on several sets of points with the same number of points and several different polygons with different numbers of sides. Then, the mean of the results is obtained.

As can be seen in Figures 4 and 5, similar to Figures 2 and 3, as the number of sides of the polygon decreases, or the number of points inside the polygon increases, the algorithm will operate more successfully due to the possibility of two connection edges between two adjacent convex pieces. In addition, the heuristic algorithm obtained the results within a shorter time than the GA provided in [18].

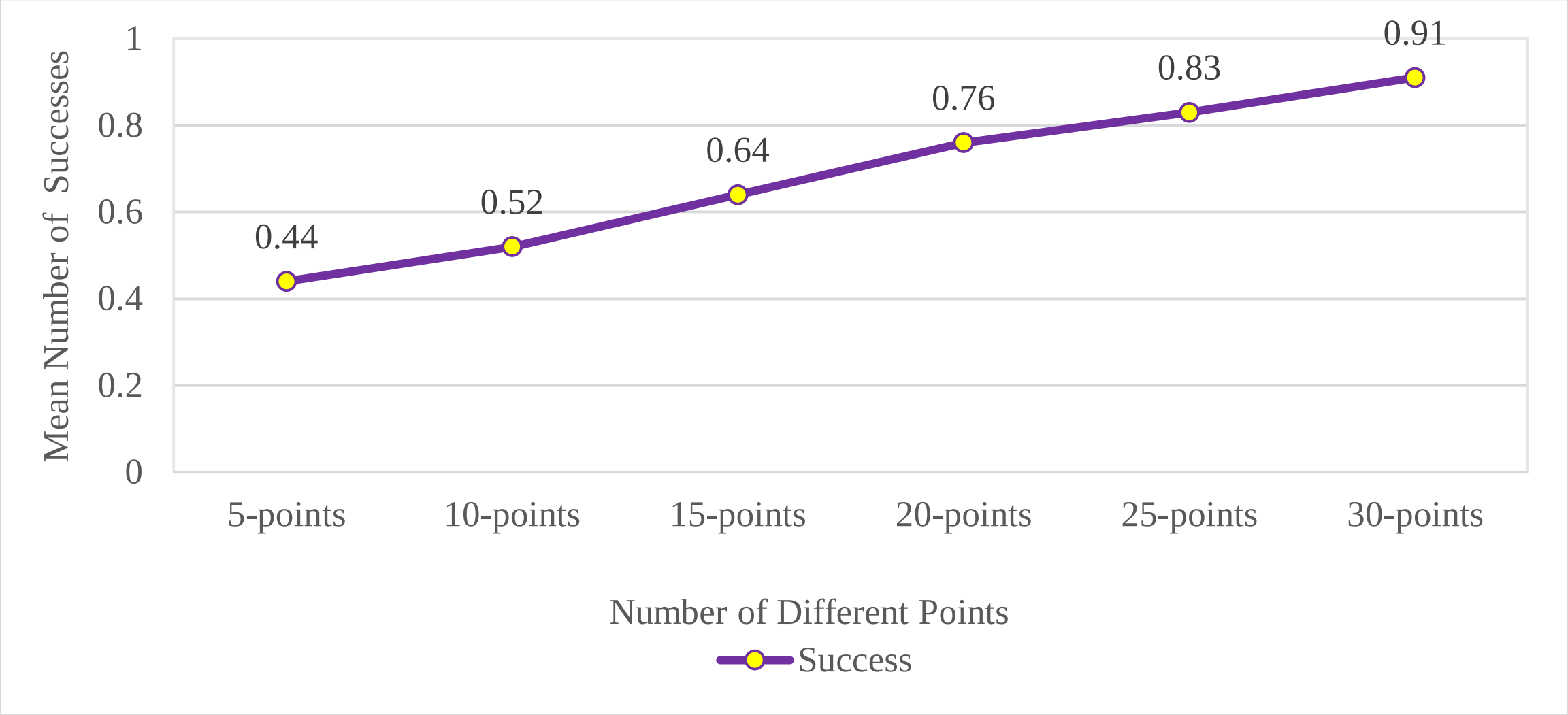

**Figure 5:** Mean Number of Successes per the Set of Points with a Different Number of Points on Different Polygons

## 4.5. Results of the Proposed Algorithm

This section provides samples of the results of the proposed algorithm. In the following figures, the red lines are the necessary diagonals for partitioning, and the blue lines are the connection edges.

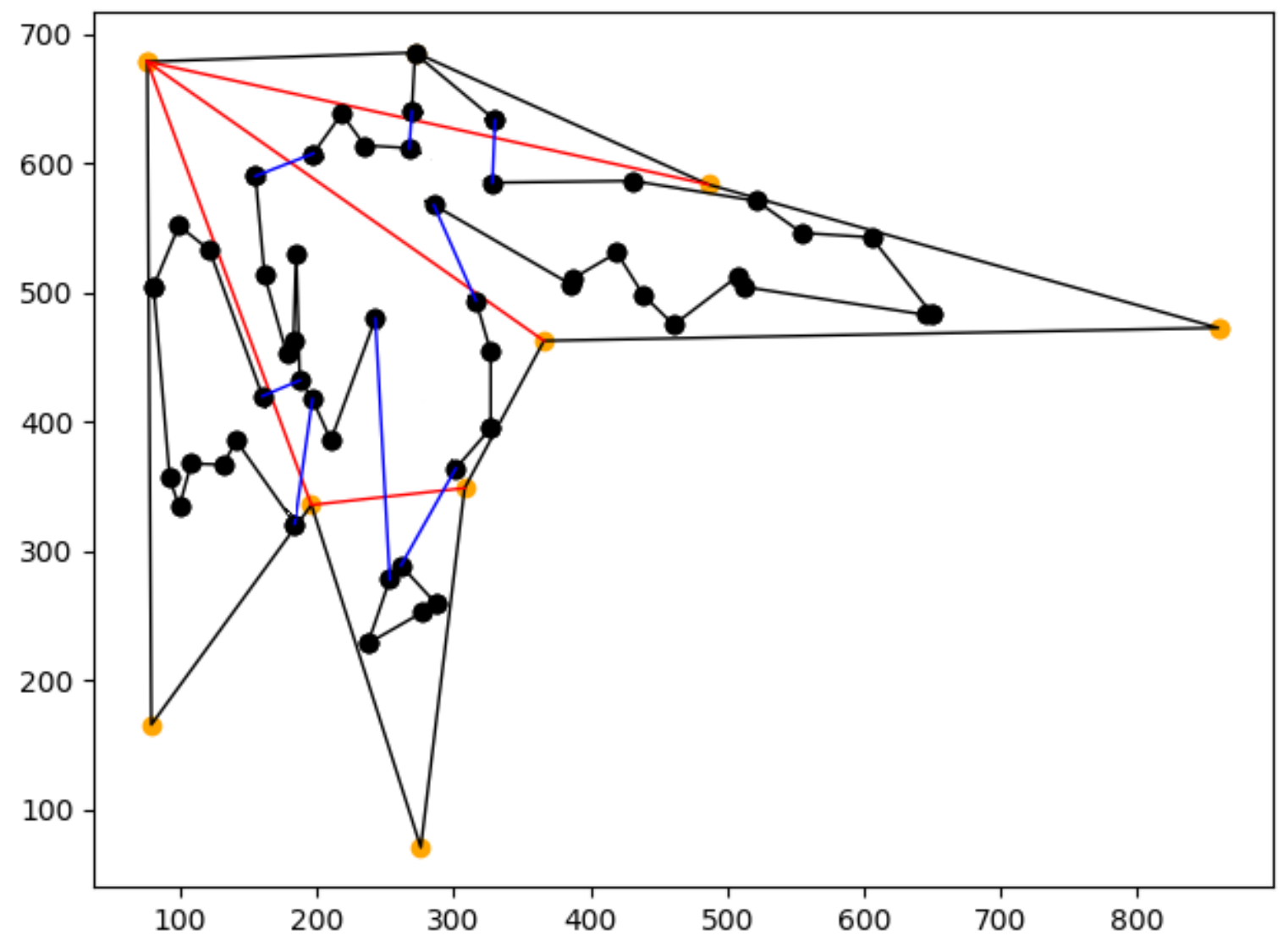

**Figure 6:** Embedded Cycle on 9-Gon with 50 Points

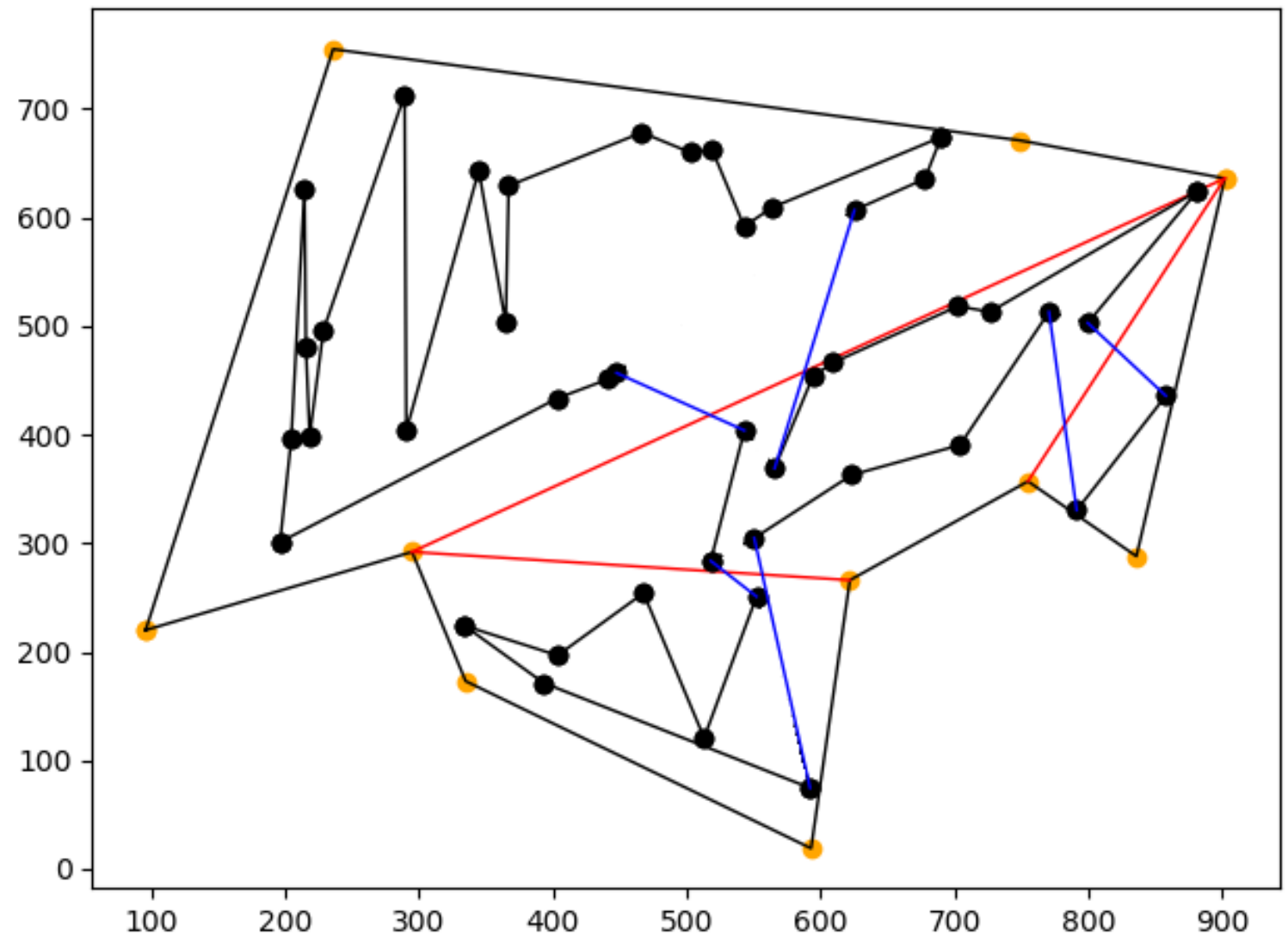

**Figure 7:** Embedded Cycle on 10-Gon with 44 Points

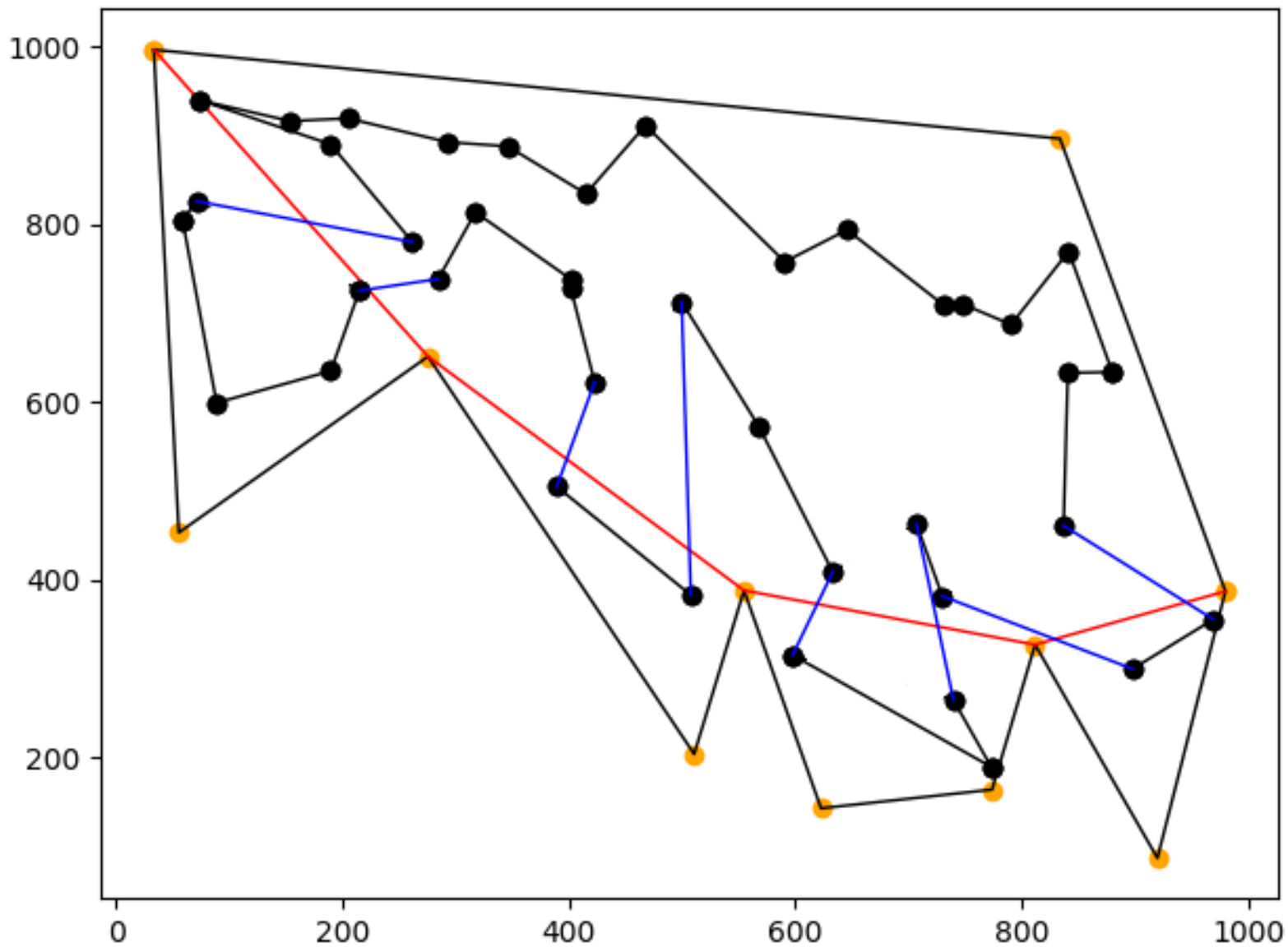

**Figure 8:** Embedded Cycle on 11-Gon with 40 Points

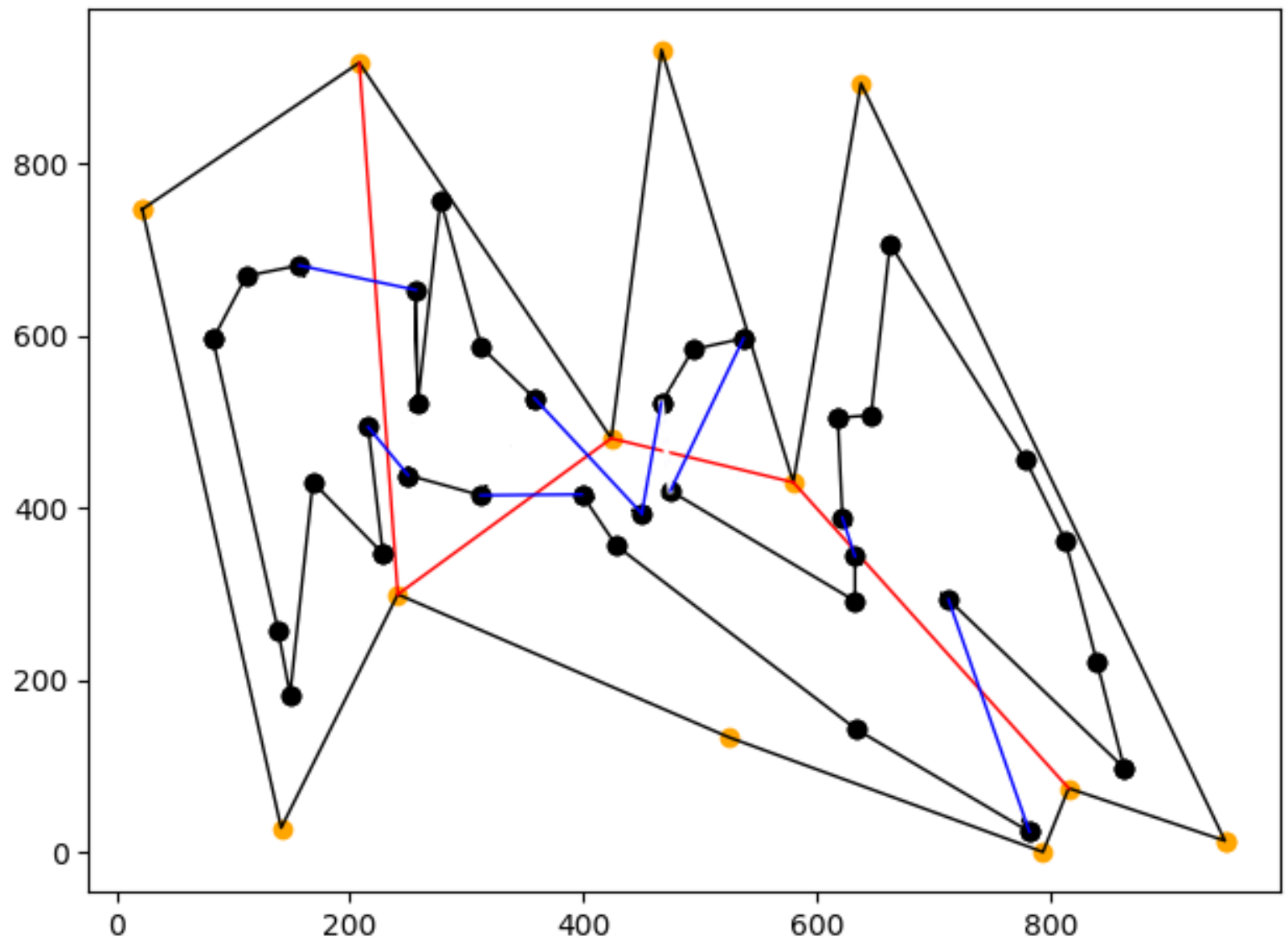

**Figure 9:** Embedded Cycle on 12-Gon with 35 Points

## 5. Conclusion and Open Problems

This paper proposed a heuristic algorithm for the straight-line planar embedding of a cycle on the points inside simple random polygons. The results of the heuristic algorithm demonstrated that as the number of points inside the polygon increases, the algorithm operated more successfully on account of the high probability of the existence of two connection edges between two adjacent convex pieces and it was able to embed a straight-line planar cycle on the points inside the polygon such that it intersects neither of the sides of the polygon.

There are several interesting open problems to our work.

1. Improvement of the heuristic algorithm proposed in this paper
2. Providing approximation algorithms with a specific approximation factor
3. Investigating the complexity of the problem proposed in this paper
4. Examining the complexity of the problem for specifying the existence or absence of a simple Hamilton (s, X, t)-path inside a simple polygon, not intersecting itself or polygon sides